\documentclass[twocolumn,printnumbers,amsmath,amssymb,showpacs,prl]{revtex4}
\usepackage{graphicx}
\usepackage{color}

\begin{document}

\title{Revealing the Link between Structural Relaxation and Dynamic Heterogeneity in Glass-Forming Liquids}

\date{\today}

\author{Lijin Wang$^{1}$}
\author{Ning Xu$^{2,\dag}$}
\author{W. H. Wang$^{3}$}
\author{Pengfei Guan$^{1,*}$}

\affiliation{$^1$Beijing Computational Science Research Center, Beijing 100193, P. R. China}

\affiliation{$^2$CAS Key Laboratory of Soft Matter Chemistry, Hefei National Laboratory for Physical Sciences at the Microscale and  Department of Physics, University of Science and Technology of China, Hefei 230026, P. R. China}

\affiliation{$^3$Institute of Physics, Chinese Academy of Sciences, Beijing 100190, P. R. China}

\begin{abstract}

Despite the use of glasses for thousands of years, the nature of the glass transition is still mysterious. On approaching the glass transition, the growth of dynamic heterogeneity has long been thought to play a key role in explaining the abrupt slowdown of structural relaxation. However, it still remains elusive whether there is an underlying link between structural relaxation and dynamic heterogeneity. Here we unravel the link by introducing a characteristic time scale hiding behind an identical dynamic heterogeneity for various model glass-forming liquids. We find that the time scale corresponds to the kinetic fragility of liquids. Moreover, it leads to scaling collapse of both the structural relaxation time and dynamic heterogeneity for all liquids studied, together with a characteristic temperature associated with the same dynamic heterogeneity.  Our findings imply that studying the glass transition from the viewpoint of dynamic heterogeneity is more informative than expected.

\end{abstract}


\maketitle

Nowadays, various theoretical and empirical equations, e.g., mode coupling (MC), Vogel-Fulcher-Tammann (VFT), Elmatad-Chandler-Garrahan (ECG), Avramov-Milchev (AM), and Mauro-Yue-Ellison-Gupta-Allan (MYEGA) forms \cite{1,2,3,4,AM,MYEGA}, are proposed to fit the structural relaxation time of glass-forming liquids, $\tau(T)$, as a function of temperature $T$, and to interpret the glass transition in different theoretical frameworks. Despite the diversity of the fitting functions, the kinetic fragility \cite{5}, $m=\partial(log \tau)/\partial(T_g/T)|_{T=T_g}$ with $T_g$ being the  glass transition temperature, is commonly employed to evaluate the deviation of $\tau(T)$ from the Arrhenius behavior \cite{1}, which proposes a useful classification of liquids along a `strong' to `fragile' scale \cite{5}. Thus, the scaling collapse of discrete $\tau(T)$ data in various glass-forming liquids is believed to be an effective way to simplify the elusive glass transition \cite{1,2,3}. Although great efforts have been devoted \cite{6,7,8,9}, it is still unclear whether there is a general and simple description (without introducing adjustable free parameters) of $\tau(T)$ for glass-forming liquids with vastly different $m$.

In the past decades, one grail in the study of glasses is the finding of dynamic heterogeneity referring to the spatiotemporal fluctuations in local dynamics \cite{10,11,12}. The growth of the dynamic heterogeneity and its dynamic correlation length \cite{13,14,15} as $T$ decreases towards the glass transition provides a possible approach to understand the dramatic slowdown of dynamics during vitrification. Thus, more attentions \cite{16,17,18,19,20,21,berthier_physics,Paluch_2012PRB,Paluch_2013JPCL,31,Coslovich_jcp} have been attracted to investigate the correlation between structural relaxation and dynamic heterogeneity in glass-forming liquids. The critical issue nowadays is that experimental and numerical studies \cite{17,18,19,20,Paluch_2013JPCL,Coslovich_jcp} have showed that dynamic heterogeneities in state points under isochronal condition (i.e., constant $\tau$) can be either invariant or variant. Recently, attempts have been made to search for the general relation between structural relaxation and dynamic heterogeneity \cite{16,berthier_physics,31}, but there seems to be no consensus on it \cite{18,Paluch_2012PRB,Paluch_2013JPCL,21}. Furthermore, the concept of fragility is believed to be correlated well with dynamic heterogeneity in  model glass-forming liquids \cite{22,23}, while an experimental study \cite{13} reported that there was no convincing correlation  between them. To our knowledge, even in model glass formers where a precise quantification of dynamic heterogeneity is feasible, the directly quantitative evidence for the correlation between fragility and dynamic heterogeneity is still lacking.

In this Letter, we reveal the underlying connection between dynamic heterogeneity and structural relaxation by introducing a characteristic time scale hidden in state points with an identical dynamic heterogeneity in different model glass-forming liquids. This time scale  corresponds to the kinetic fragility of glass-forming liquids and bridges structural relaxation and dynamic heterogeneity by achieving fantastic scaling collapses. Moreover, a rather general description of $\tau(T)$ for various glass-forming liquids can be achieved from the viewpoint of constant dynamic heterogeneity condition without introducing any free parameter.

We perform extensive molecular dynamics simulations in the $NPT$ (constant number of particles $N$, pressure $P$, and temperature $T$) ensemble in six potential models \cite{24}: harmonic (Harm), Hertzian (Hertz), 12-6 repulsive Lennard-Jones (RLJ), 36-6 RLJ, 12-6 Lennard-Jones (LJ), and embedded atom method (EAM) potentials.We measure the self-part of the intermediate scattering function \cite{25}, $F_s(k,t)=\frac{1}{N}\langle \sum_{j=1}^{N} {\rm exp}\{{\rm i}\vec{k}\cdot[\vec{r}_{j}(t)-\vec{r}_{j}(0)]\}\rangle$, where $\vec{r}_j(t)$ is the location of particle $j$ at time $t$, $|\vec{k}|$ takes approximately the value at the first peak of the static structure factor \cite{26}, and $\langle.\rangle$ denotes time average. The structural relaxation time $\tau$ is defined by the relation \cite{24}: $F_s(k,\tau)=e^{-1}$. Dynamic heterogeneity  is  quantified by the time-dependent non-Gaussian parameter \cite{10,21,27}, $\alpha_2(t)=\frac{3}{5}\langle\Delta {r}^4\rangle/\langle\Delta {r}^2\rangle^2-1$, where $\Delta {r}$ is the  displacement of a particle during time  $t$. In Supplemental Material (SM) \cite{24}, we also show results regarding dynamic heterogeneity characterized by the four-point dynamic susceptibility $\chi_4$ \cite{15,28,Coslovich_jcp}.

\begin{figure}
\includegraphics[width=0.48\textwidth]{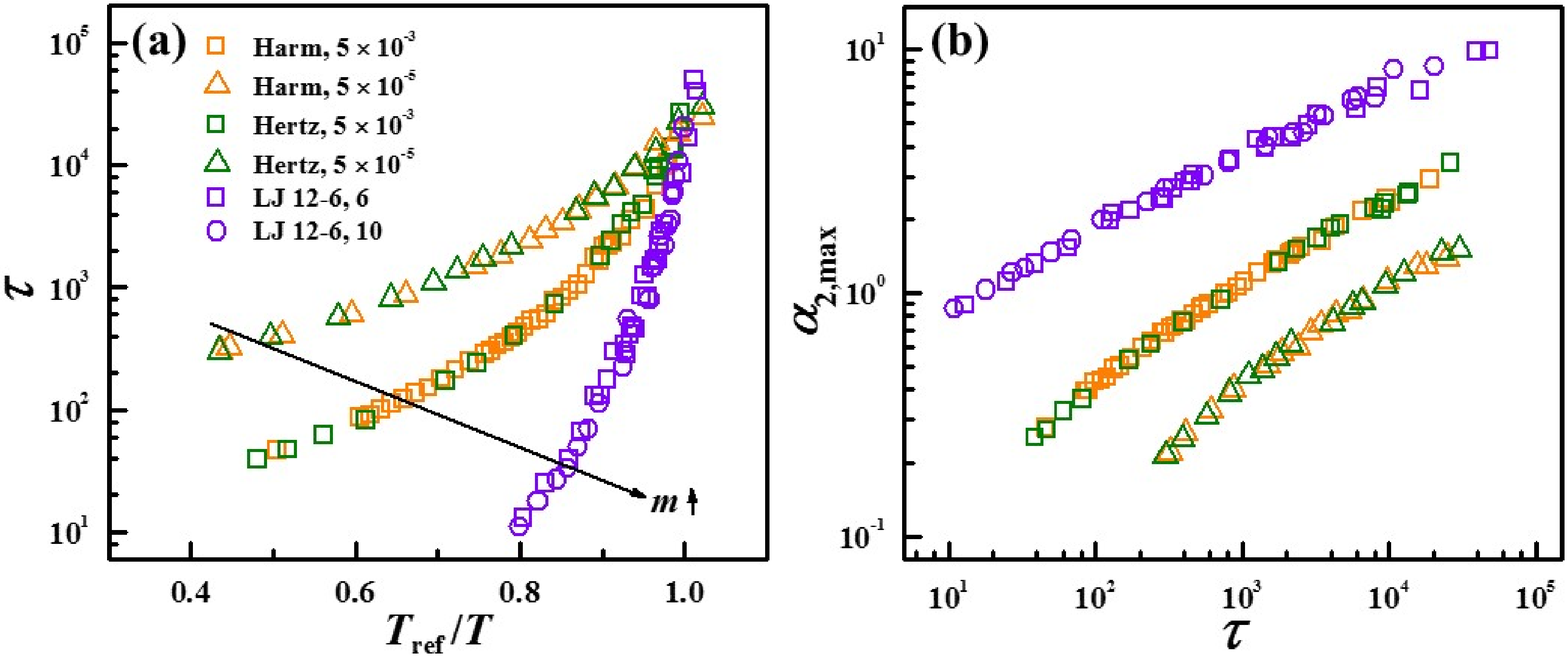}
\caption{\label{fig:fig1} (color online) (a) Angell plots of structural relaxation time $\tau$ versus scaled reciprocal temperature $T_{ref}/T$ for systems with different potentials and pressures. $T_{ref}$ is determined according to  $\tau(T_{ref} )=\tau_g \approx 2.16 \times 10^4$.  (b) Correlation between $\tau$ and $\alpha_{2,max}$. Symbols in panels (a) and (b) have the same meanings.}
\end{figure}

Figure \ref{fig:fig1}(a) shows the Angell plots of $\tau$ versus $T_{ref}/T$ for six glass-forming systems with different potentials and pressures. $T_{ref}$ is a reference temperature at which $\tau=\tau_g$ is sufficiently large in the endurable time window of simulation and identical for all systems, which is treated here as $T_g$ to calculate the kinetic fragility $m$. Two systems have identical $m$ if their curves in Fig. \ref{fig:fig1}(a) coincide, and steeper curve represents a more fragile liquid with a larger $m$. Systems with different potentials could exhibit the same $m$, as illustrated by the collapse of curves with harmonic and Hertzian potentials at the same pressures. With increasing pressure, the kinetic fragility increases, consistent with previous simulation studies \cite{22,29,wangJCP}, whereas  the pressure dependence of fragility in most real liquids \cite{Paluch_jcp_fragility,Grzybowska_jpclandjcp_fragility,Roland_prb_fragility} is different from model ones. Therefore, by varying the pressures and potentials, we are able to investigate systems with vastly different values of $m$  \cite{24} ( the range of $m$ is still not as large as that in real materials \cite{Paluch_jcp_fragility,Grzybowska_jpclandjcp_fragility,Roland_prb_fragility}).

 On approaching the glass transition, $\alpha_2(t)$ exhibits a non-monotonic $t$ dependence with a maximum $\alpha_{2,max}$ occurring at $t=\tau_{\alpha_{2,max}} $ (see examples in Fig. S2 in  SM \cite{24}). As expected \cite{10,21,27}, both $\alpha_{2,max}$ and $\tau_{\alpha_{2,max}}$ increase when $T$ decreases. Figure \ref{fig:fig1}(b) shows the correlation between $\alpha_{2,max}$  and $\tau$ for the same systems shown in Fig. \ref{fig:fig1}(a). For each system, $\alpha_{2,max}$ increases with increasing $\tau$, indicating that dynamic heterogeneity grows with the slowdown of structural relaxation during vitrification \cite{12,13,14,15}. Moreover, along with Fig. \ref{fig:fig1}(a), Fig. \ref{fig:fig1}(b) shows that, under the isochronal condition, one system with a larger $m$ exhibits a larger $\alpha_{2,max}$. Therefore, more fragile liquids are more heterogeneous in dynamics \cite{22,23}. More importantly, systems with the same $m$ also exhibit identical $\alpha_{2,max}(\tau)$, which implies that the kinetic fragility is very likely to be the long-sought key parameter to connect structural relaxation and dynamic heterogeneity.

\begin{figure}
\includegraphics[width=0.5\textwidth]{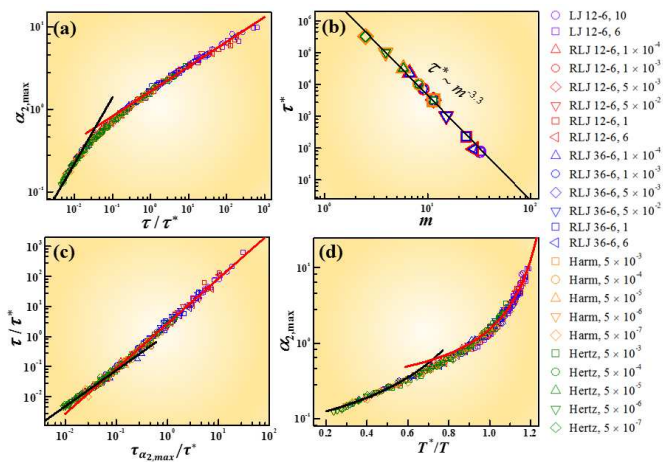}
\caption{\label{fig:fig2} (color online). (a) Maximum non-Gaussian parameter $\alpha_{2,max}$ versus  reduced structural relaxation time $\tau / \tau^*$ with $\tau^*$ being the characteristic time scale under the iso-$\alpha_{2,max}$ condition ($\alpha_{2,max} \approx 1.67$ here). The solid lines are fits to $\alpha_{2,max}  \sim (\tau/\tau^* )^{\nu}$ with $\nu=0.8$ (black line) and $\nu=0.3$ (red line).  (b) Correlation between $\tau^*$ and kinetic fragility $m$. $m$ is calculated at $T_g=T_{ref}$ as set in Fig. \ref{fig:fig1}(a). The black solid line is a fit to $\tau^*  \sim m^{-\gamma}$, where $\gamma=3.3$. (c) Universal scaling between $\tau$ and $\tau_{\alpha_2,max}$. The solid lines are fits to
$\tau/\tau^* \sim (\tau_{\alpha_{2,max}}/\tau^*)^\beta$, where $\beta=1.2$ (black line) and $\beta=1.5$ (red line). (d) $\alpha_{2,max}$ versus scaled reciprocal temperature $T^*/T$ with $T^*$ being the characteristic temperature. The black and red solid lines are fitting curves consistent with the VFT fitting in Fig. \ref{fig:fig3} and power-law fittings in Fig. \ref{fig:fig2}(a) [Eq.(\ref{eq4}) can be derived from Eqs. (\ref{eq1}) and (\ref{eq5})]: $\alpha_{2,max}=0.093  \emph{exp}[1.285/(T/T^*-0.705)]$ and $\alpha_{2,max}=0.299 \emph{exp} [0.482/(T/T^*-0.705)]$, respectively. }
\end{figure}

Figure \ref{fig:fig2}(a) shows that we can collapse the $\alpha_{2,max}(\tau)$ curves for all systems investigated  onto a single master curve when $\tau$ is scaled by $\tau^*$, and hence
  \begin{equation}
\alpha_{2,max}=f_{\tau}(\tau/\tau^*), \label{eq1}
\end{equation}
where $f_{\tau}$ is piecewise [see fitting lines in Fig. \ref{fig:fig2}(a)]. Here, the scaling parameter $\tau^*$ is a system-dependent characteristic time scale for all systems to have the same $\alpha_{2,max}$, i.e., under the iso-$\alpha_{2,max}$ condition. We choose a Hertzian system at $T=1.46\times10^{-4}$ and $P=5.00\times10^{-3}$ as a reference state, for which $\tau^*\approx 3.25 \times 10^3$ and $\alpha_{2,max}\approx 1.67$. The scaling collapse is obtained by shifting all other curves onto that of the reference state. Surprisingly, Fig. \ref{fig:fig2}(b) shows that $\tau^* \sim m^{-\gamma}$, so Eq. (\ref{eq1}) can be rewritten as
  \begin{equation}
\alpha_{2,max}=f_{m}(\tau m^{\gamma}), \label{eq2}
\end{equation}
where $\gamma$ varies with the time $\tau_g$ used to evaluate $m$. As shown in Figs. \ref{fig:fig1}(a) and \ref{fig:fig2}(b), here we choose $\tau_g \approx 2.16\times 10^4$, and $\gamma \approx 3.3$ and $m$ takes values from $2.5$ to $32.4$ (in Fig. S3 of  SM \cite{24}, we show another example with $\gamma \approx 4.5$ and $m \in [7.9,54.6]$, when $\tau_g = 10^6$  is chosen).  Recent studies \cite{17,18,19,20,Coslovich_jcp,Paluch_2013JPCL} showed that whether dynamic heterogeneities at constant $\tau$  vary  depends on different control parameters, e.g.,  the softness of atomic interactions \cite{17}, pressures \cite{18,20} and density scaling \cite{Coslovich_jcp,Paluch_2013JPCL}. However, to our knowledge, it  remains unknown whether there is a single parameter that controls the  correlation between dynamical heterogeneity and $\tau$ in different glass formers. Our findings in Figs. \ref{fig:fig1} and \ref{fig:fig2} suggest that it is the kinetic fragility that couples with the characteristic time scale hiding behind the iso-$\alpha_{2,max}$ condition and plays a key role in establishing the general relation between $\alpha_{2,max}$ and $\tau$. Thus, Equation (2) reveals quantitatively the underlying correlation between dynamic heterogeneity and structural relaxation in glass-forming liquids.

 It is  interesting to compare experimental results with ours. Some experimental studies \cite{13,Roland_pre} reported no convincing (or weak) correlation between fragility and dynamic heterogeneity  from an indirect measure of $\chi_4$, which is in contrary to the obvious correlation shown in Fig. \ref{fig:fig2}(b).  Further studies are thus required to resolve the disagreement between experiments of real materials and simulations of model glass formers and  to examine the generality of the correlation between fragility and dynamic heterogeneity observed here.

Although both $\tau$ and $\tau_{\alpha_{2,max}}$  increase upon cooling, they are usually not linearly related, i.e., the characteristic times for structural relaxation and  establishment of $\alpha_{2,max}$ decouple \cite{21,27}. As shown in Fig. S4 of  SM \cite{24}, the $\tau$ versus $\tau_{\alpha_{2,max}}$  curves for systems with different fragilities deviate a lot.  A scaling collapse of $\tau(\tau_{\alpha_{2,max}})$ has been achieved by simply adjusting a system-dependent scaling factor to  rescale $\tau_{\alpha_{2,max}}$ \cite{21}. However, the physical meaning of the manipulative scaling factor is unclear. Interestingly, when we plot $\tau / \tau^*$ against $\tau_{\alpha_{2,max}}/\tau^*$, as shown in Fig. \ref{fig:fig2}(c), curves for all systems studied collapse onto the same master curve:
  \begin{equation}
\tau / \tau^*=H(\tau_{\alpha_{2,max}}/\tau^*), \label{eq3}
\end{equation}
where $H$ is also piecewise. Since $\tau^*$ is intrinsically equivalent to $m$, the decoupling relation between $\tau$ and $\tau_{\alpha_{2,max}}$ is controlled as well by the kinetic fragility, which is another robust evidence confirming that the kinetic fragility is the key to connecting structural relaxation and dynamic heterogeneity.

Now we have seen the essential role of the kinetic fragility or the characteristic time scale  in unifying the relationship between structural relaxation and dynamic heterogeneity. This further stimulates our ambition to find a general description of $\tau (T)$. Note that $\tau^*$  hides behind an identical $\alpha_{2,max}$, which couples with a system-dependent temperature $T^*$. Now that we have shown the importance of $\tau^*$, it's interesting to know whether $T^*$ is crucial as well.

Like what has been done for Fig. \ref{fig:fig2}(a), we shift all $\alpha_{2,max} (T)$ curves (see  examples in Fig. S5 in  SM \cite{24}) to that of the Hertzian one at $P=5.00\times10^{-3}$ and take the Hertzian state at $T=T^*=1.46 \times 10^{-4}$ and $P=5.00\times10^{-3}$ as the reference. This leads to a nice scaling collapse:
  \begin{equation}
\alpha_{2,max}=f_{T}(T^*/T), \label{eq4}
\end{equation}
as shown in Fig. \ref{fig:fig2}(d). Equation (\ref{eq4}) verifies that $T^*$ is indeed the characteristic temperature we are looking for. Like $f_\tau$ in Eq. (\ref{eq1}) and $H$ in Eq. (\ref{eq3}), $f_T$ in Eq. (\ref{eq4}) is piecewise as well. Similarly,  piecewise behaviours can also be observed when $\chi_{4,max}$ (the maximum of $\chi_4$) is plotted as a function of $\tau$ or $T$ \cite{berthier_physics,31,24}. The initial power-law and then a logarithmic growth of $\chi_{4,max}$  with $\tau$ can be predicted, respectively, by mode coupling and random first order transition theories  though details regarding the crossover between the two regimes of growth are still puzzling \cite{12}.  Since $\alpha_{2,max}$ grows less strongly than $\chi_{4,max}$ with $\tau$ or $T$ (see Fig.~S10 in SM \cite{24}),  it may be interesting to check whether theories that can predict behaviours of $\chi_{4,max}(\tau)$ are also applicable to  $\alpha_{2,max}(\tau)$.

Unlike that $\tau^*$ has a one-to-one correspondence with $m$, we find no direct correlation between $T^*$ and $m$, e.g., two systems with LJ 12-6 potential at $P=6$ and $P=10$, respectively, have equal $m$ but pretty different values of $T^*$  \cite{24}. Equation (4) hints that although $T^*$ is sensitive to system parameters (interaction potentials, pressures, etc.), it may be coupled to other characteristic temperatures, e.g., the glass transition temperature, which is crucial to establish the general description of $\tau (T)$ for various systems shown in the following.

\begin{figure}
\includegraphics[width=0.48\textwidth]{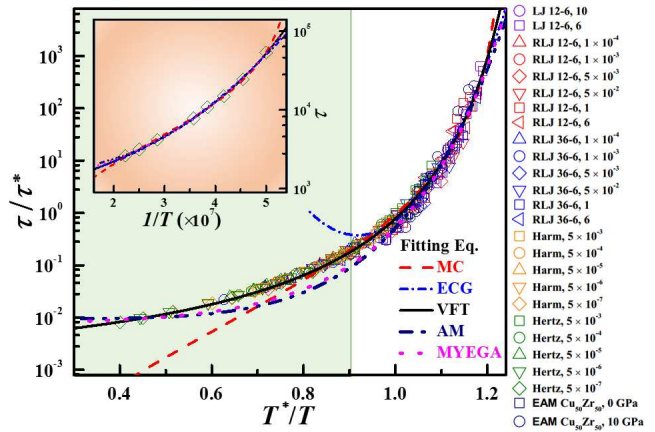}
\caption{\label{fig:fig3} (color online). Main panel: Scaled structural relaxation time, $\tau/\tau^*$, versus scaled reciprocal temperature, $T^*/T$, for all systems studied. Black solid curve is the VFT fit: $y=0.00337\emph{exp}[1.606/(x-0.705)]$, where $x=T/T^*$ and $y=\tau/\tau^*$. Red dashed curve is a fit to the MC form: $y=0.00314 (x-0.809)^{-3.441}$. Blue dash-dot curve indicates the ECG fit: $y=0.373 \emph{exp}[95.845(x^{-1}-0.917)^2 ]$. Navy dash-dot-dot curve is a fit to the AM form: $y=0.00943 \emph{exp} [(1.284/x)^{5.542}]$. Magenta dotted curve is the MYEGA fit: $y=0.00782 \emph{exp} [(0.0730/x)\emph{exp}(4.052/x)]$ .
Inset: $\tau(T)$ for a Hertz system at $P=5\times 10^{-7}$, whose corresponding scaled data lie in the shadowed region in the main panel. Note that the curve in the inset can be fitted well with all the above five forms before scaling. After scaling, it lies in the region where only VFT works in the main panel.}
\end{figure}

Now we are going to move one step further to discuss the scaling collapse of $\tau(T)$. To unify in the same framework different dynamic slowdown in various glass-forming liquids, people have tried to manipulate the scaling collapse of dynamics in different ways.  An excellent scaling collapse of $\tau(T)$ in Lennard-Jones systems has been achieved by using a density scaling function \cite{6}. However, the density scaling procedure usually yields different scaling curves for different systems and fails in some systems, e.g., systems with harmonic  potentials studied here. Moreover, it has been shown that existing methods to achieve scaling collapse of $\tau(T)$ for specific systems cannot be simply generalized to other systems \cite{7,8,9}.
Interestingly, the combination of Eqs. (\ref{eq1}) and (\ref{eq4}) can lead to the general scaling relation of $\tau(T)$:
\begin{equation}
\tau / \tau^*=f_{\tau}^{-1} [f_{T}(T^*/T)]=F(T^*/T), \label{eq5}
\end{equation}
Therefore, by introducing $\tau^*$ and $T^*$, the long-sought scaling collapse of $\tau(T)$ for various systems is straightforward, as corroborated in Fig. \ref{fig:fig3}. To our knowledge, so far, there has been no work to successfully collapse $\tau(T)$ for so many  systems with vastly different potentials and over so wide a range of pressures and fragilities without introducing additional or arbitrary parameters. The scaling collapse shown in Fig. \ref{fig:fig3} only involves characteristic scales  associated with an identical dynamic heterogeneity, which have clear physical meanings. Dynamic heterogeneity is believed to be important in understanding the glass transition, which is directly and confirmatively evidenced here by the scaling collapses shown in Figs. \ref{fig:fig2} and \ref{fig:fig3}.

Next, we study the functional form for $F(x)$ in Eq. (\ref{eq5}). As mentioned earlier, there are multiple functions proposed to fit $\tau(T)$. In the inset to Fig. \ref{fig:fig3}, we show that VFT, MC,  ECG, AM and MYEGA forms can all fit $\tau(T)$ well for a single system. However, when we try to fit the master curve in the main panel of Fig. \ref{fig:fig3} using these five forms, only VFT can fit the whole curve nicely, while the other four forms can only fit the high $T^*/T$ part well, which mainly contains more fragile liquids within the simulation time window. Though the VFT form can describe our master curve well, it should also be noted that the VFT description of $\tau(T)$ is challenged in an experimental study \cite{HecksherNP} where ultraviscious molecular liquids were studied.

Our numerical studies of different model glass formers unravel a general description of the dynamics during vitrification. For the underlying connection between dynamic heterogeneity and structural relaxation, the key is the awareness of the importance of constant dynamic heterogeneity condition and the characteristic scales hiding behind it. By introducing the characteristic time scale   and temperature  under the iso-$\alpha_{2,max}$  condition, scaling collapses regarding the structural relaxation and dynamic heterogeneity can be generally described. It reveals the long-sought general description of the relationship between structural relaxation time and temperature without introducing any adjustable parameter in various glass formers. Since the characteristic time scale is equivalent to kinetic fragility, it is suggested that the kinetic fragility serves as the link between structural relaxation and dynamic heterogeneity. Moreover, our major conclusions   hold as well if dynamic heterogeneity is quantified by dynamic susceptibility, as discussed systematically in  SM \cite{24}. Our work suggests that dynamic heterogeneity plays a more important role than expected in studying the nature of the glass transition.

Though our major findings do not rely on whether dynamic heterogeneity is measured by $\alpha_{2,max}$ or  $\chi_{4,max}$, it should  be  noted that $\alpha_{2,max}$ and $\chi_{4,max}$ (see  comparison between $\alpha_{2,max}$ and $\chi_{4,max}$ in Fig.~S10 of SM \cite{24}) as well as the size of cooperatively rearranging regions proposed in Adam-Gibbs model \cite{30} are only qualitatively equivalent measures of dynamic heterogeneity, because quantitative inconsistencies of their temperature dependence  can be observed \cite{jcp_difference}. This can also to some extent be  implied by the observation that $\tau_{\alpha_{2,max}}(\tau)$ is piecewise  while $\tau_{\chi_{4,max}} \sim \tau$ \cite{13,Coslovich_jcp} (see Figs.~S4 and S6(b) of SM \cite{24}) with  $\tau_{\chi_{4,max}}$ being the time when $\chi_{4,max}$ occurs.

Our findings here are based on numerical studies of molecular glass formers, and the glass transition temperature defined here is higher than experiment \cite{3,31,32}, which thus call for further experimental verification of our findings. To our knowledge, the development of an  estimate of $\chi_{4,max}$ in ref.\cite{13} has advanced greatly  the experimental studies of dynamic heterogeneity in real materials. Since $\alpha_{2,max}$ is also important in measuring dynamic heterogeneity from our study, it will  be meaningful to devise an experimentally measurable estimate of $\alpha_{2,max}$. Probing  dynamic heterogeneity precisely is challenging in experiments of molecular glass formers \cite{12,18,20,13,31} while it is feasible in experiments of colloids \cite{33,34,35}. Recent studies \cite{8,33} have demonstrated that some behaviours of colloidal and molecular glass formers show remarkable similarities, and hence it's intriguing to see whether the scenarios reported here can also be observed in colloidal experiments.

We wish to thank Y. Hu, B. Shang, M. D. Shattuck and C. S. O'Hern for helpful discussions. We also acknowledge the computational support from the Beijing Computational Science Research Center (CSRC). L. W. and P. G. are supported by the National Natural Science Foundation of China (Grant No. 51571011), the MOST 973 Program (Grant No. 2015CB856800) and the NSAF joint program (Grant No. U1530401). N. X. is supported by the National Natural Science Foundation of China (Grants No. 11734014 and No. 11574278) and Fundamental Research Funds for the Central Universities (Grant No. 2030020028).

\end{document}